\documentclass[iop,apjl]{emulateapj}
\usepackage{amssymb}		 
\usepackage{amsmath, amsthm, amssymb,amsfonts}
\usepackage[english]{babel}
\usepackage{epstopdf}
\usepackage{color}
\usepackage{lipsum}
\usepackage{graphicx}

\newcommand{\rmd}{{\rm d}}
\newcommand{\pD}[2]{\frac{\partial #2}{\partial #1}}
\newcommand\bb[1]{\mbox{\boldmath{$#1$}}}
\newcommand\grad{\bb{\nabla}}
\newcommand\bcdot{\,\bb{\cdot}\,}
\newcommand\btimes{\,\bb{\times}\,}
\newcommand{\mc}[1]{\mathcal{#1}}
\newcommand\bs[1]{\boldsymbol{#1}}
\newcommand{\ez}{\hat{\bb{z}}}
\newcommand{\eb}{\hat{\bb{b}}}
\newcommand{\vth}[1]{v_{{\rm th}{\rm {#1}}}}
\newcommand{\FM}{F_{\rm M}}

\begin{document}

\title{Dual Phase-space cascades in 3D hybrid-Vlasov--Maxwell turbulence}
\author{S.~S.~Cerri$^1$}
\author{M.~W.~Kunz$^{1,2}$}
\author{F.~Califano$^3$}
\affiliation{$^1$Department of Astrophysical Sciences, Princeton University, Princeton, NJ 08544, USA}
\affiliation{$^2$Princeton Plasma Physics Laboratory, PO Box 451, Princeton, NJ 08543, USA}
\affiliation{$^3$Dipartimento di Fisica ``E. Fermi'', Universit\`a di Pisa, 56127 Pisa, Italy}
\email{E-mail of corresponding author: scerri@astro.princeton.edu}

\begin{abstract}
To explain energy dissipation via turbulence in collisionless, magnetized plasmas, the existence of a dual real- and velocity-space cascade of ion-entropy fluctuations below the ion gyroradius has been proposed. Such a dual cascade, predicted by the gyrokinetic theory, has previously been observed in gyrokinetic simulations of two-dimensional, electrostatic turbulence. For the first time we show evidence for a dual phase-space cascade of ion-entropy fluctuations in a three-dimensional simulation of hybrid-kinetic, electromagnetic turbulence. Some of the scalings observed in the energy spectra are consistent with a generalized theory for the cascade that accounts for the spectral anisotropy of critically balanced, intermittent, sub-ion-Larmor-scale fluctuations. The observed velocity-space cascade is also anisotropic with respect to the magnetic-field direction, with linear phase mixing along magnetic-field lines proceeding mainly at spatial scales above the ion gyroradius and nonlinear phase mixing across magnetic-field lines proceeding at perpendicular scales below the ion gyroradius. Such phase-space anisotropy could be sought in heliospheric and magnetospheric data of solar-wind turbulence and has far-reaching implications for the dissipation of turbulence in weakly collisional astrophysical plasmas.
\end{abstract}


\maketitle

\section{Introduction}\label{sec:intro}

Turbulence is one means by which a fluid or plasma nonlinearly transforms kinetic and/or electromagnetic energy into thermodynamic heat. In a fluid or collisional plasma, this conversion is mediated by molecular viscosity and/or resistivity, which dissipatively remove power from small spatial scales. However, many astrophysical and space plasmas are so hot and diffuse that their collisional mean free paths are as large as, if not bigger than, the macroscopic scales of interest. 
Examples include the solar wind, the intracluster medium, and low-luminosity accretion flows.
In this situation, the production of small-scale (``kinetic'') structure in velocity space can compensate for the small collisionality and eventually lead to collisional relaxation and irreversible heating. In other words, turbulent collisionless plasmas find a nonlinear route to dissipation through phase space \citep[see][]{SchekochihinPPCF2008}.

One way to generate this velocity-space structure is via {\em linear phase mixing}, caused by the ballistic response of the particle distribution function and associated with \citet{Landau1946} damping. In a magnetized plasma, this process occurs primarily along magnetic-field lines, being stifled across them by the smallness of particles' gyroradii. 
Another route to generating small-scale structure in velocity space is {\em nonlinear phase mixing}, by which the particles' distribution function at a given position is nonlinearly mixed by decorrelated ring-averaged fluctuations. This mechanism, predicted by the gyrokinetic theory \citep{DorlandHammettPFB1993,SchekochihinPPCF2008,SchekochihinAPJS2009,PlunkJFM2010}, causes a dual cascade in both the real and velocity spaces perpendicular to the magnetic field of fluctuations in the ion distribution function---the so-called {\em entropy cascade}.

Classical turbulence theory is formulated in Fourier space, where nonlinear interactions between fluctuations produce a flux of energy to larger wavenumbers with varying degrees of wavevector anisotropy with respect to the magnetic-field direction \citep[e.g.,][]{Iroshnikov1963,Kraichnan1965,GoldreichSridhar1995,Boldyrev2006}. An analogous approach can be used to study velocity-space cascades by instead employing Hermite polynomials or Hankel functions as the basis \citep[e.g.,][and references therein]{PlunkJFM2010,AdkinsSchekochihinJPP2017}.

Unfortunately, velocity-space cascades have traditionally been difficult to diagnose, both in satellite data and in numerical simulations. Although several 3D simulations of kinetic/gyrokinetic turbulence exist \citep[e.g.,][]{HowesPRL2008,HowesPRL2011,ServidioJPP2015,ToldPRL2015,WanPOP2016,CerriAPJL2017,Groselj2017,FranciAPJ2018,Arzamasskiy2018}, thus far the velocity-space cascade produced by nonlinear phase mixing has been directly diagnosed only in 2D simulations of gyrokinetic electrostatic turbulence \citep{TatsunoPRL2009,Tatsuno12}, with little other indirect evidence \citep{BanonNavarroPRL2011,CerriPOP2014a}. Numerical evidence of linear phase mixing and its associated velocity-space cascade exists, but still within reduced settings \citep[e.g.,][]{WatanabeSugamaPOP2004,HatchJPP2014,ParkerPOP2016,GroseljAPJ2017}. Observationally, only recent advances in spacecraft instrumentation have provided the first evidence of a velocity-space cascade (in the electron distribution function) occurring in the solar wind filling the Earth's magnetosheath \citep{ServidioPRL2017}.

In this Letter, we report on the occurrence of a phase-space cascade in a high-resolution 3D-3V simulation of hybrid-Vlasov--Maxwell (HVM) turbulence in a collisionless plasma with finite electron inertia. We also derive scaling laws for the ion-entropy cascade, generalized to account for different spectral anisotropies and a possible reconnection-mediated energy transfer. Some of these scalings are in agreement with our simulation results.

\section{Sub-Ion-Larmor Turbulence}\label{sec:entropycascade}

\subsection{Kinetic-Alfv\'{e}n-wave and ion-entropy cascades}

As turbulent energy arrives from the ``MHD'' inertial range at $k_\perp \rho_i \sim 1$ ($k_\perp$ is the wavenumber perpendicular to the magnetic field, $\rho_i$ is the ion Larmor radius), it can be resonantly absorbed and/or redistributed into various phase-space cascades. 
Two examples of the latter are those of kinetic Alfv\'en waves (KAWs) and of ion-entropy fluctuations \citep{SchekochihinPPCF2008,SchekochihinAPJS2009,Kunz2017}. 
By way of reviewing their predicted turbulence scaling laws, we generalize them to account for different wavevector anisotropies.

We measure ion-entropy fluctuations via
\begin{equation} \label{eq:dEfull-def}
\delta{\cal E}\equiv
-\int\rmd^3\bb{v} \, \bigl(f\ln f - \FM \ln \FM \bigr), 
\end{equation}
where $\FM=\FM(t,\bb{r},\bb{v})$ is an isotropic Maxwellian whose number density $n$, mean flow $\bb{u}$, and temperature $T$ matches, respectively, the zeroth, first, and second moments of the ion distribution function $f=f(t,\bb{r},\bb{v})$. Often, a linearized version of Equation (\ref{eq:dEfull-def}) is defined, $\delta{\cal E}\approx\int\rmd^3\bb{v}\, (T\delta f^2/2\FM)$ with $\delta f\equiv f-\FM$, which features in a collisionless invariant called the gyrokinetic free energy \citep{SchekochihinAPJS2009}. For low-frequency, sub-ion-Larmor-scale KAW turbulence, this $\delta f$ is asymptotically equal to the non-adiabatic ``gyrokinetic'' response $h$, i.e., $\delta f\approx h$ and $\delta{\cal E}\approx \int\rmd^3\bb{v} \, (T h^2/2\FM)$ for $k_\perp \rho_i \gg 1$. 
In the gyrokinetic theory, the KAW and ion-entropy cascades are energetically decoupled, with $h$ being a passive tracer of the ring-averaged KAW turbulence in phase space. 

As the fluctuations cascade via nonlinear interactions to $k_\perp \rho_i \gg 1$, particles with different gyroradii but similar guiding-center positions---i.e., with different velocities perpendicular to the magnetic field---experience different fluctuations. Plasma particles belonging to different portions of $f$ then undergo different turbulent evolution, leading to phase mixing in the velocity space perpendicular to the magnetic field ($v_\perp$) with correlation scale
\begin{equation}\label{eq:dvprp}
\frac{\delta v_\perp}{\vth{i}} \sim \frac{1}{\rho_i} \left| \frac{v_\perp}{\Omega_i} - \frac{v'_\perp}{\Omega_i} \right| \sim \frac{1}{k_\perp \rho_i} \ll 1 ,
\end{equation}
where $\vth{i} \equiv \sqrt{2T/m_{\rm i}}$ is the ion thermal speed, $m_{\rm i}$ is the ion mass, and $\Omega_{\rm i}$ is the ion Larmor frequency (see \citealt{SchekochihinAPJS2009}, \S 7.9.1). In the Hermite representation of velocity space, $\delta v_\perp / \vth{i} \sim m^{-1/2}_\perp$, where $m_\perp$ is the order of the Hermite basis function, and so Equation (\ref{eq:dvprp}) implies $m_\perp \propto k^2_\perp$. 
The real- and velocity-space cascades of ion-entropy fluctuations are thus tightly entwined and occur simultaneously: ion-entropy fluctuations are mixed in velocity space by decorrelated fluctuations in real space, and the resulting decorrelated velocity-space structure impacts the ring-averaging of those fluctuations. Scaling laws for the entropy cascade should reflect this mixing. 

In \S\ref{subsec:entropycascade-aniso} we derive such scaling laws using arguments borrowed from gyrokinetic theory. In doing so, we are neither promoting the gyrokinetic theory as a generic description of sub-ion-Larmor-scale turbulence, nor are we claiming that our simulation parameters (see \S\ref{sec:entropycascade-hvm}) are best suited to test such a theory, which is based on a low-frequency, small-amplitude, spatially anisotropic asymptotic ordering. Rather, the agreement or lack thereof between the consequent scaling theory and our simulation results highlights the degree to which certain aspects of the gyrokinetic theory manifest in a more general (Vlasov) setting. It is important to note, however, that the concept of correlated real- and velocity-space fluctuations in the sub-ion-Larmor range (as in Equation (\ref{eq:dvprp})) does not require especially low frequencies; it is a generic consequence of particles with similar guiding centers but gyroradii differing by an amount $\sim$$k^{-1}_\perp$ experiencing decorrelated electromagnetic fluctuations.

\subsection{Spectral anisotropy and turbulence scaling laws}\label{subsec:entropycascade-aniso}

We denote the scales perpendicular and parallel to the magnetic-field direction by $\lambda$ and $\ell_\parallel$, respectively. The relation between these scales is parametrized as follows:
\begin{equation} \label{eq:spectr-anisotropy}
\ell_{\|,\lambda}\propto\lambda^{\alpha/3},
\end{equation}
where $\alpha$ describes the spectral anisotropy of the fluctuations in the sub-ion-Larmor range of interest. $\alpha=1$ returns the ``standard'' scalings for KAW turbulence \citep[e.g.][]{SchekochihinAPJS2009,BoldyrevAPJ2013};
$\alpha=2$ corresponds to the intermittency-corrected scenario proposed by \citet{BoldyrevPerezAPJL2012}; 
and $\alpha=3$ describes an ``isotropic'' cascade, which has been measured in some hybrid-PIC simulations \citep{FranciAPJ2018,Arzamasskiy2018}.

Using the characteristic timescale of linear KAWs ($\tau_{\rm KAW,\lambda} \propto \lambda \ell_{\parallel,\lambda}$) and assuming a critically balanced cascade in which the scale-dependent linear and nonlinear timescales are comparable, the nonlinear timescale $\tau^{(\alpha)}_{\rm nl,\lambda}$ at perpendicular scale $\lambda$ satisfies 
\begin{equation} \label{eq:t_nl-t_kaw}
\tau_{\rm nl,\lambda}\sim\tau_{\rm KAW,\lambda}^{(\alpha)}\propto \lambda^{1+\alpha/3}.
\end{equation}
Following the arguments in \citet[][\S 7.9.2]{SchekochihinAPJS2009}, the nonlinear timescale in the entropy cascade is obtained by weighting $\tau^{(\alpha)}_{\rm KAW,\lambda}$ by the factor $(\rho_i/\lambda)^{1/2}$ due to the ring averaging in the gyrokinetic nonlinearity: 
\begin{equation} \label{eq:tkaw-ringavg}
\widetilde{\tau}_{h,\lambda} \sim \left(\frac{\rho_i}{\lambda}\right)^{1/2}\tau_{\rm KAW,\lambda}^{(\alpha)}.
\end{equation}
With $\lambda \ll \rho_i$, during each KAW correlation time the nonlinearity changes the scale-dependent gyrokinetic response $h_\lambda$ only by a small factor, $\Delta h_{\lambda}/h_{\lambda} \sim \tau_{\rm KAW,\lambda}^{(\alpha)}/\widetilde{\tau}_{h,\lambda} \ll 1$. These changes accumulate as a random walk, i.e., as $(t/\tau_{\rm KAW,\lambda}^{(\alpha)})^{1/2} \Delta h_{\lambda}$. The entropy cascade time $\tau_{h,\lambda}^{(\alpha)}$ is the time needed to produce an order-unity change in $h_{\lambda}$, i.e., $(\tau_{h,\lambda}^{(\alpha)}/\tau_{\rm KAW,\lambda}^{(\alpha)})^{1/2}\Delta h_{\lambda}\sim h_{\lambda}$. Thus,
\begin{equation} \label{eq:tau_h}
\tau_{h,\lambda}^{(\alpha)} \sim \left(\frac{\rho_i}{\lambda}\right) \tau_{\rm KAW,\lambda}^{(\alpha)} \sim \lambda^{\alpha/3} .
\end{equation}
Assuming a constant entropy flux through scales, $h_{\lambda}^2/\tau_{h,\lambda}\sim\varepsilon_h = {\rm const}$, we obtain
\begin{equation} \label{eq:h_lambda}
h_{\lambda}^{(\alpha)} \propto \lambda^{\alpha/6},
\end{equation}
which corresponds to the following spectra of $h$ and $\delta{\cal E}$:
\begin{gather} \label{eq:h-spectrum_alpha}
E_h(k_\perp) \propto k_\perp^{-(3+\alpha)/3}, \\
\label{eq:dEntr-spectrum_alpha}
E_{\delta{\cal E}}(k_\perp)\propto k_\perp^{-(3+2\alpha)/3} .
\end{gather}
For standard KAW-turbulence anisotropy $k_\|\propto k_\perp^{1/3}$ ($\alpha=1$), the predicted $h$ spectrum $\propto$$k_\perp^{-4/3}$ is recovered \citep{SchekochihinAPJS2009}, and $E_{\delta{\cal E}} \propto k_\perp^{-5/3}$. In the intermittency-corrected case ($\alpha=2$), $k_\|\propto k_\perp^{2/3}$ and so $E_h \propto k_\perp^{-5/3}$ and $E_{\delta{\cal E}}\propto k_\perp^{-7/3}$. For isotropic sub-ion-scale turbulence ($\alpha=3$), $k_\|\propto k_\perp$ and so $E_h \propto k_\perp^{-2}$ and $E_{\delta{\cal E}} \propto k_\perp^{-3}$. 
In all of these cases, the corresponding spectra in parallel wavenumber are independent of $\alpha$:
\begin{equation} \label{eq:h-spectrum_kpara}
E_h(k_\|) \propto k_\|^{-2} 
\quad{\rm and}\quad
E_{\delta{\cal E}}(k_\|) \propto k_\|^{-3} .
\end{equation}

Accompanying these real-space spectra are velocity-space spectra described by the perpendicular Hermite number $m_\perp$. Using $m_\perp \propto k^2_\perp$ (see Equation (\ref{eq:dvprp})) in Equation (\ref{eq:h-spectrum_alpha}) implies $E_h(m_\perp) \propto m^{-(6+\alpha)/6}_\perp$. Note that $E_h(m_\perp) \propto m^{-7/6}_\perp$ for $\alpha = 1$, which is equivalent to the $E_h(p) \propto p^{-4/3}$ Hankel spectrum predicted by \citet{PlunkJFM2010} for nonlinear phase mixing in 2D electrostatic gyrokinetic turbulence (note that $p \sim \sqrt{m_\perp}$). In the intermittency-corrected case, $E_h(m_\perp) \propto m^{-4/3}_\perp$.

We caution that some of these scalings may be degenerate with a turbulent cascade whose energy transfer is mediated by magnetic reconnection \citep[e.g.][]{CerriCalifanoNJP2017,FranciAPJL2017}. As the cascade proceeds toward smaller spatial scales and the wavevector anisotropy increases, the reconnection timescale of the turbulent fluctuations may eventually become comparable to the nonlinear cascade time \citep{LoureiroBoldyrevAPJ2017,MalletJPP2017}. As discussed in \citet{MalletJPP2017}, this process will naturally ``reset'' the wavevector anisotropy and can thus provide an alternative explanation for spectral anisotropies with $\alpha>1$. To explore this possibility, we adopt the scale-dependent reconnection timescale as the nonlinear cascade time for the turbulent cascade upon which the ion-entropy cascade develops; that is, we substitute $\tau_{\rm rec,\lambda}$ for $\tau_{\rm KAW,\lambda}$ in the arguments leading to Equations (\ref{eq:h-spectrum_alpha}) and (\ref{eq:dEntr-spectrum_alpha}). Using the rates given in \citet{LoureiroBoldyrevAPJ2017} for $\beta\sim1$, {\em viz.}~$\tau_{\rm rec,\lambda}^{(n)} \propto \lambda^{(4n+2)/(3n)}$ (the index $1 < n\leq 2$ parametrizes cases where the tearing-mode parameter $\Delta' \sim \lambda^{-1} (k\lambda)^{-n}$, with $\lambda$ being a proxy for the current-sheet thickness), it is easy to show that $h_\lambda^{(n)}\propto\lambda^{(n+2)/(6n)}$. For the limiting cases $n=1$ and $n=2$, we recover the spectra in Equations (\ref{eq:h-spectrum_alpha}) and (\ref{eq:dEntr-spectrum_alpha}) for $\alpha=3$ and $\alpha=2$, respectively. For $n=1$, the reconnection rate, $\propto$$\lambda^{-2}$, at kinetic scales becomes large enough to efficiently disrupt the turbulent ``eddies'', producing an effectively isotropic cascade ($\alpha=3$). For $n=2$, such disruption events may occur, but perhaps not so efficiently as to completely isotropize the cascade; sheet-like structures may then persist long enough to produce the intermittency-corrected spectral anisotropy, $\alpha=2$.

\section{Ion-entropy cascade in HVM turbulence}\label{sec:entropycascade-hvm}

The possibility of a dual phase-space cascade at sub-ion-Larmor scales is investigated using a nonlinear simulation of decaying turbulence with the Eulerian (i.e., grid-based both in real and velocity space) HVM code \citep{ValentiniJCP2007}.
The model equations governing the ion distribution function $f(t,\bb{r},\bb{v})$ and the electromagnetic fields $\bb{E}(t,\bb{r})$ and $\bb{B}(t,\bb{r})$ are the Vlasov equation, Faraday's law of induction, and a generalized Ohm's law that assumes quasi-neutrality and includes the inductive and Hall electric fields, a thermo-electric field driven by pressure gradients in the (isothermal) electron fluid, and the leading-order electron inertia term~\citep{CerriAPJL2017}:
\begin{equation}\label{eq:HVM_Vlasov}
\pD{t}{f} + \bb{v}\bcdot\grad f + \bigl( \bb{E} + \bb{v}\btimes\bb{B} \bigr) \bcdot \pD{\bb{v}}{f} = 0,
\end{equation}
\begin{equation}\label{eq:HVM_Ohm}
\bigl(1-d_{\rm e}^2\nabla_\perp^2\bigr) \bb{E} = -\bb{u}\btimes\bb{B} + \frac{(\grad\btimes\bb{B})\btimes\bb{B}}{n} - \frac{T_{\rm e} \grad n}{n},
\end{equation}
\begin{equation}\label{eq:HVM_Maxwell}
\pD{t}{\bb{B}} = -\grad\btimes\bb{E} ,
\end{equation}
where all quantities are normalized using $m_{\rm i}$, $\Omega_{\rm i}$, the Alfv\'{e}n speed $v_{\rm A} \equiv B / \sqrt{4\pi m_{\rm i} n}$, and the ion inertial length $d_{\rm i} \equiv v_{\rm A}/\Omega_{\rm i}$. The electron inertia term in Equation (\ref{eq:HVM_Ohm}) involving the electron inertial length $d_{\rm e}$ is the only term able to physically break flux freezing and allow magnetic reconnection to occur. To correctly capture such physics, a reduced mass ratio $m_{\rm i}/m_{\rm e} = 100$ has been adopted so that $d_{\rm e} = 0.1d_{\rm i}$ is spatially resolved.

The simulation (published in \citealt{CerriAPJL2017}) was initialized with a stationary, spatially homogeneous, Maxwellian, ion-electron plasma, threaded by a uniform magnetic field $\bb{B}_0 = B_0 \ez$ and characterized by a plasma beta parameter $\beta_{{\rm i}0} = \beta_{{\rm e}0} \equiv 8\pi n_0 T_{0} / B^2_0 = 1$ (the subscript ``0'' denotes an initial value). Large-scale, random, nearly isotropic 3D magnetic perturbations $\delta\bb{B}$ are placed on top of $\bb{B}_0$, with $0.1\leq(k\rho_{\rm i})_{\delta\bs{B}} \leq 0.5$ and $\delta B_{\rm rms}\simeq 0.23$. The ion phase space is discretized using $(N_x,N_y,N_z,N_{v_x},N_{v_y},N_{v_z}) = (384,384,64,51,51,51)$ uniformly distributed points. Periodic boundary conditions are imposed in real space, with $L_z \simeq 62 \rho_{\rm i}$ and $L_x = L_y \simeq 31 \rho_{\rm i}$, corresponding to a spectral domain spanning $0.2\leq k_{x,y} \rho_{\rm i}\leq 38.4$ and $0.1\leq k_z \rho_{\rm i}\leq 3.2$. Velocity space is limited in each direction by $v_{\rm max} = \pm 5\sqrt{T_0/m_{\rm i}}$, beyond which $f=0$.

The initial $\delta\bb{B}$ freely decays into fully developed turbulence after a few Alfv\'en-crossing times, when a peak in the rms current density and quasi-stationary energy spectra are obtained. Our analysis is performed during this fully developed turbulent state, which is characterized by intermittent magnetic-field fluctuations across and below $\rho_{\rm i}$ and by a spectral anisotropy that, in the $k_\perp \rho_{\rm i} \gtrsim 1$ kinetic range, is consistent with this intermittency, {\it viz.},~$k_\|\propto k_\perp^{2/3}$.

\begin{figure}
\centering
 \includegraphics[width=0.48\textwidth]{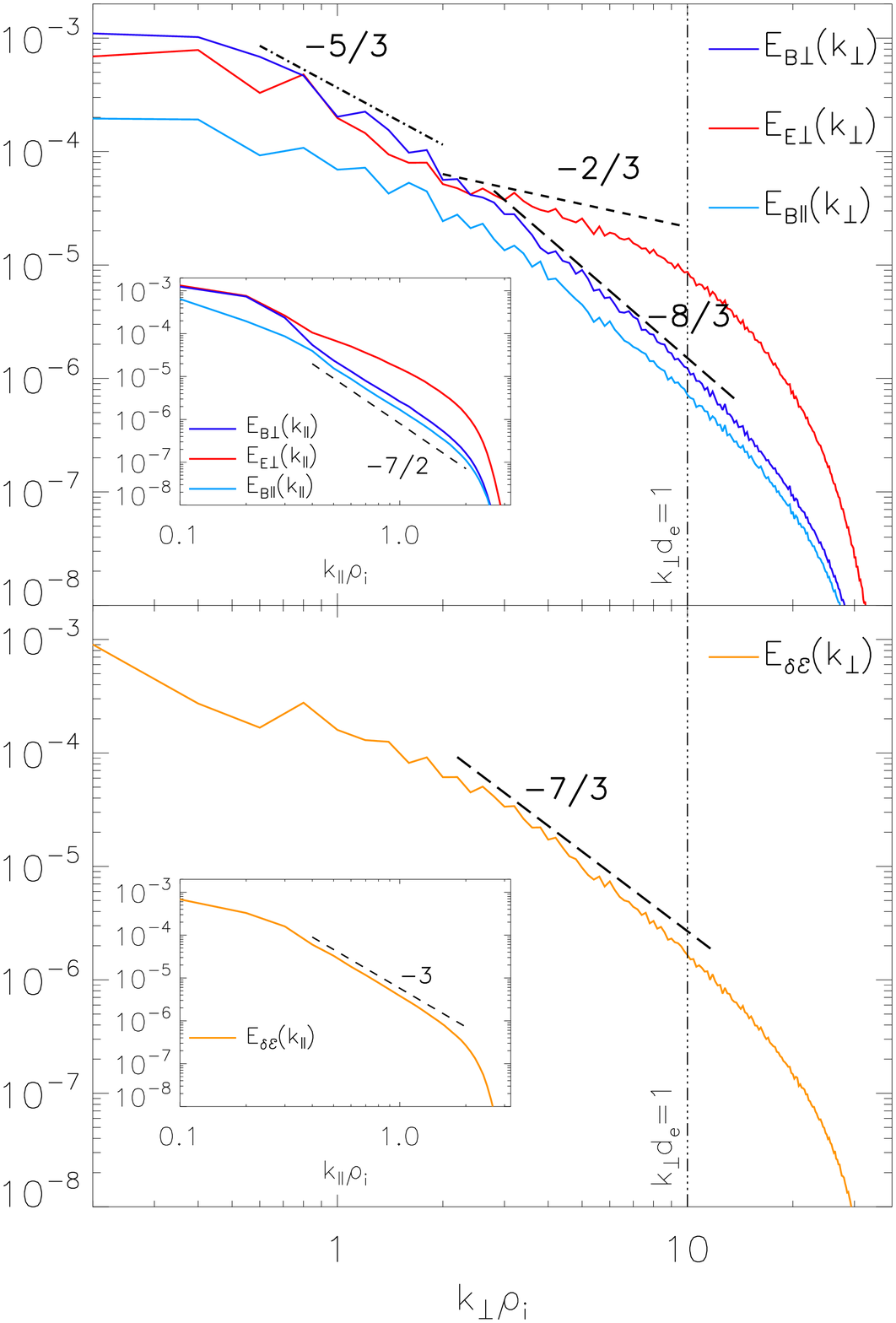}
 \caption{Energy spectra of the perpendicular electric field $E_{E\perp}$ (red solid line), the perpendicular magnetic field $E_{B\perp}$ (blue solid line), the parallel magnetic-field fluctuations $E_{B\|}$ (light-blue solid line), and the non-thermal entropy fluctuations $E_{\delta{\cal E}}$ (orange solid line), all versus $k_\perp$. Insets report the same spectra versus $k_\|$. Labelled dashed and dot-dashed lines provide reference slopes.}
 \label{fig:EBEntrSpect_kperp}
\end{figure} 

\subsection{Real-space cascades}

The top panel of Figure \ref{fig:EBEntrSpect_kperp} displays energy spectra of the perpendicular electric field $E_{E\perp}$ (red solid line), the perpendicular magnetic field $E_{B\perp}$ (blue solid line), and the parallel magnetic-field fluctuations $E_{B\|}$ (light-blue solid line) versus $k_\perp$. The inset shows the same spectra versus $k_\|$. (The parallel electric-field spectrum, not shown, is at least an order of magnitude smaller than any other spectrum.) Here, ``$\|$'' and ``$\perp$'' are defined with respect to $\bb{B}_0$. Both $E_{B\perp}$ and $E_{B\|}$ exhibit a break near $k_\perp \rho_i \approx 2$, with a sub-ion-Larmor-scale slope close to $-8/3$. The perpendicular electric-field spectrum at these scales is shallower, with slope initially close to $-2/3$ but progressively steepened by electron inertial effects. The spectral cutoffs at $k_\perp \rho_{\rm i}\gtrsim 20$ and $k_\parallel \rho_{\rm i}\gtrsim 2$ are caused by electron inertia and (weak) spectral filters.

The bottom panel of Figure \ref{fig:EBEntrSpect_kperp} displays $E_{\delta{\cal E}}(k_\perp)$; the inset shows $E_{\delta{\cal E}}(k_\|)$. In the kinetic range, below the scales at which the magnetic-field spectrum breaks, the $\delta{\cal E}$ spectrum shows a spectral slope near $-7/3$. The corresponding $k_\|$ spectrum has a slope very close to $-3$. Both are in remarkable agreement with the theory presented in \S\ref{subsec:entropycascade-aniso} for $\alpha=2$, derived by treating $\delta{\cal E}$ as a passive scalar of intermittent KAW turbulence.

\subsection{Phase-space representation of the cascade}

The ion-entropy cascade is predicted to be a dual cascade simultaneously in real and velocity space. To test this idea, we must Fourier--Hermite transform $\delta f$. However, due to computational expense, the full 3D-3V distribution function spanning the entire computational domain cannot be analyzed at once. Instead, we have analyzed two reduced quantities separately: (i) a $v_z$-integrated $\delta f$ measured at $8$ different $z$ locations, $\langle \delta f(x,y,z=\{z_0\}_{j=1\dots 8},v_x,v_y)\rangle_{v_z}$; and (ii) a ($v_x$,$v_y$)-integrated $\delta f$ with no spatial information removed, $\langle\delta f(x,y,z,v_z)\rangle_{v_x,v_y}$. These quantities are Fourier--Hermite transformed using a maximum Hermite mode number $M=30$ (accounting for our finite velocity-space resolution). The first reduced quantity then gives the spectrum in the ($k_\perp$,$m_\perp$) plane, while the second is representative of the ($k_\perp$,$k_\|$,$m_\|$) space. Because of the Fourier transform, here ``$\|$'' and ``$\perp$'' are defined with respect to $\bb{B}_0$, so that $m_\| = m_z$ and $m_\perp = (m^2_x+m^2_y)^{1/2}$.

\begin{figure}
\centering
\includegraphics[width=0.48\textwidth]{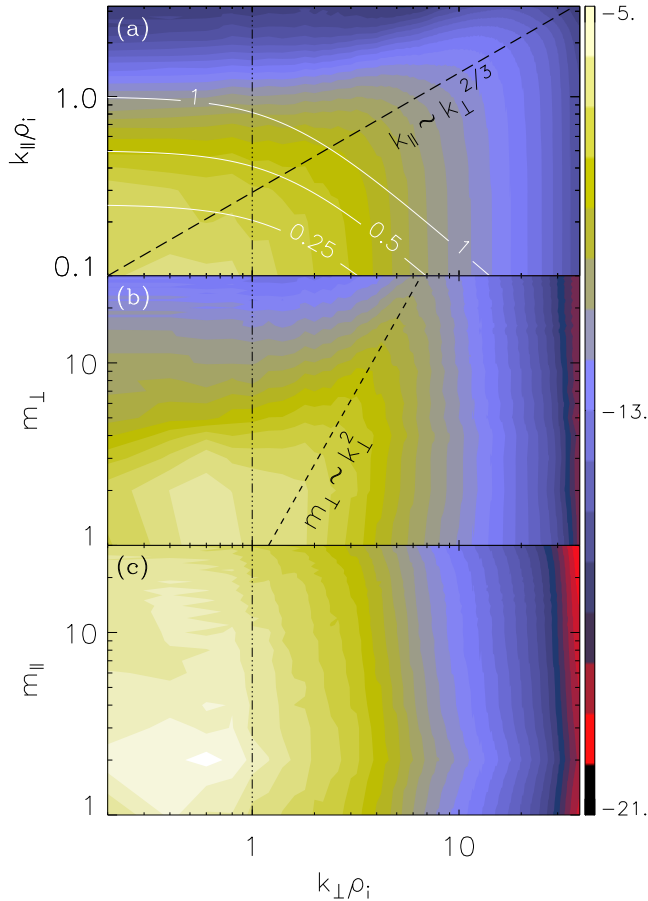}
\caption{Fourier--Hermite spectrum of the (reduced) non-thermal distribution function, $\log|\delta f_{{\bs k}, {\bs m}}|^2$. Top to bottom: ($k_\|$,$k_\perp$), ($m_\perp$,$k_\perp$), and ($m_\|$,$k_\perp$) planes. Labelled dashed lines provide reference slopes. White solid lines in panel (a) trace $\omega_{\rm KAW}/\Omega_{\rm i}=0.25$, $0.5$, and $1$ isocontours.}
 \label{fig:df_km}
\end{figure}

Figure \ref{fig:df_km} displays contour plots of these reduced phase-space spectra in the ($k_\perp$,$k_\|$), ($k_\perp$,$m_\perp$), and ($k_\perp$,$m_\|$) planes. We observe several interesting features. The entropy cascade in real space develops a spectral anisotropy consistent with $k_\|\propto k_\perp^{2/3}$ (panel a), the same as occurs in the electromagnetic fields \citep{CerriAPJL2017}. 
Note that within the hybrid-kinetic model and in the wavenumber range explored by our simulation, effects ordered out of gyrokinetics (e.g., finite Larmor frequency) may be present. Indeed, the white solid lines in Figure \ref{fig:df_km}(a), which trace $\omega_{\rm KAW}/\Omega_{\rm i}=0.25$, $0.5$, and $1$ isocontours, indicate that the linear KAW frequency $\omega_{\rm KAW}$ can exceed $\Omega_{\rm i}$ in our simulation.\footnote{For simplicity, we have used the approximate expression $\omega_{\rm KAW} = k_\parallel v_{\rm A}\sqrt{1 + 0.5 (k_\perp \rho_{\rm i})^2}$ for $\beta_{\rm i} = T_{\rm i}/T_{\rm e} = 1$, which matches the result from the linear gyrokinetic theory to within $\approx$2\% in the wavenumber regime of interest.} 
Nevertheless, perpendicular phase mixing occurs at $k_\perp\rho_i\gtrsim 1$ and consists of a dual phase-space cascade, occuring approximately along $m_\perp \propto k^2_\perp$ (panel b). A similar calculation of the Fourier--Hankel spectrum of $\delta f$ (not shown) reveals $p\propto k_\perp$, as predicted by \citet{PlunkJFM2010} and consistent with $m_\perp \propto k^2_\perp$. (Using a gyro-averaged $\delta f$ does not qualitatively change these results.) Parallel phase mixing occurs simultaneously, but is confined mainly to $k_\perp\rho_i\lesssim 1$ (panel c) and $k_\|\rho_i\lesssim 0.5$ (not shown), in agreement with the conjecture by \citet[][\S 7.9.4]{SchekochihinAPJS2009} that nonlinear perpendicular phase mixing is more efficient than linear parallel phase mixing at sub-ion-Larmor scales.

\subsection{Velocity-space cascades}

We now examine the velocity-space spectrum of $\delta f$. We extracted $\delta f(x,y,z,v_x,v_y,v_z)$ from $8$ reduced local sub-domains of spatial size $\ell_x=\ell_y\simeq\rho_i$ and $\ell_z\simeq 7\rho_i$, each consisting of $12\times12\times8$ grid points, and then performed a 3D Hermite transform at each grid point. Using the {\em local} magnetic-field direction $\eb\equiv\bb{B}/B$, the resulting spectra were transformed into a field-aligned coordinate system ($m_{\perp 1}$,$m_{\perp 2}$,$m_\|$) with $m_\perp = (m^2_{\perp 1} + m^2_{\perp 2})^{1/2}$. 

\begin{figure}
\centering
 \includegraphics[width=0.48\textwidth]{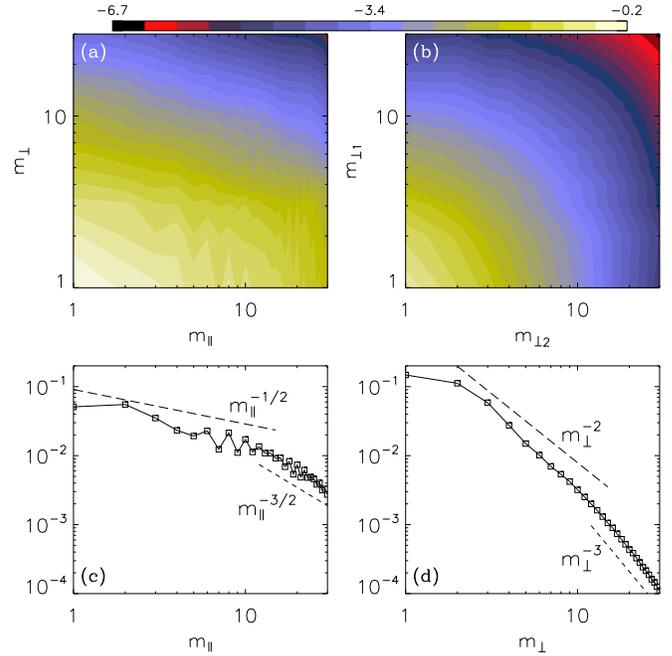}
\caption{Spatially averaged Hermite spectrum of the non-thermal distribution function, $\log|\delta f_{\bf m}|^2$, in a {\em local} magnetic-field-aligned coordinate system $(m_\|,m_\perp)$ with $m_\perp = (m^2_{\perp 1} + m^2_{\perp 2})^{1/2}$. Labelled dashed lines provide reference slopes.}
\label{fig:2DPlots}
\end{figure}

The resulting (spatially averaged) Hermite spectrum of $\delta f$ is shown in Figure \ref{fig:2DPlots}.\footnote{Despite $\delta f$ having no zeroth, first, or second moment, the  Hermite spectra exhibit residual $m=1$ and $m=2$ coefficients. There are two causes of this: (i) a finite velocity-space resolution is used when subtracting off the local, shifted Maxwellian $\FM$ from $f$ and then projecting $\delta f$ onto the discrete Hermite basis; and (ii) $\FM$ is taken to be isotropic in velocity space, even though magnetic-field-biased temperature anisotropy $T_\| \neq T_\perp \neq T$ is observed.}  
The spectrum is clearly anisotropic and gyrotropic with respect to $\eb$ (panels a and b), as expected for magnetized turbulence \citep[e.g.][]{ServidioPRL2017}. 
The $m_\|^{-1/2}$ spectrum in Figure \ref{fig:2DPlots}(c) for $m_\| \lesssim 15$ likely reflects linear phase mixing occurring along magnetic-field lines, as predicted by  \citet{ZoccoSchekochihinPOP2011} and \citet{KanekarJPP2015} (cf.~\citealt{WatanabeSugamaPOP2004}). The $m_\perp$ spectrum, on the other hand, is noticeably steeper than predicted, being closer to $-2$ (for $m_\perp \lesssim 15$). The same $m_\perp^{-2}$ spectrum is consistently recovered when integrating the $\delta f$ spectrum in Figure \ref{fig:df_km}(b) over $k_\perp$; an analogous procedure applied to the Fourier--Hankel transform of $\delta f$ returns a compatible spectrum close to $p^{-3}$. Likewise, integrating the spectrum in Figure \ref{fig:df_km}(b) over $m_\perp$ yields a spectrum close to $k_\perp^{-3}$ in the range where the dual cascade is observed, also steeper than our prediction for $E_h(k_\perp)$ (see Equation (\ref{eq:h-spectrum_alpha})). These results hold true for the gyro-averaged $\delta f$ as well. While we have no explanation for this discrepancy, we do note that \citet{ServidioPRL2017} predict a $m^{-2}$ spectrum and $m$-space anisotropy when magnetic-field fluctuations play a dominant role in the velocity-space cascade (by contrast with the gyrokinetic case, in which the $v_\perp$ cascade is driven predominantly by electric-field fluctuations). The difference may be a result of the relatively weak spectral anisotropy in our simulation, which causes the linear KAW frequency to approach $\Omega_{\rm i}$ somewhat early in the sub-ion-Larmor range (see Figure \ref{fig:df_km}), violating the gyrokinetic ordering used in \S\ref{sec:entropycascade} to predict the $\mc{E}_{h}$ spectrum.

Both the parallel and perpendicular Hermite spectra steepen at $m_{\|,\perp} \approx 15$ to approximately $m_\|^{-3/2}$ and $m_\perp^{-3}$. Although a $-3/2$ slope is predicted by \citet{AdkinsSchekochihinJPP2017} and \citet{ServidioPRL2017} for the regime of Vlasov turbulence where advection or electric-field fluctuations dominate, because of the finite velocity-space resolution in our simulation our measured slopes in this range may not be converged.

\section{Conclusions}

Using a high-resolution HVM simulation, we have shown for the first time that a dual phase-space cascade is occurring in 3D-3V electromagnetic turbulence. Generalized scaling laws that account for different spectral anisotropies with respect to the magnetic-field direction have been derived, and the observed ion-entropy cascade is consistent with some of these new scalings in the intermittency-corrected case, $k_\|\propto k_\perp^{2/3}$ \citep{BoldyrevPerezAPJL2012}. The non-thermal distribution function $\delta f$ is shown to develop both real- and velocity-space structure that is anisotropic with respect to the magnetic-field direction. Both the real-space cascades of $\delta f$ and the electromagnetic fields exhibit a sub-ion-Larmor-scale spectral anisotropy consistent with $k_\|\propto k_\perp^{2/3}$. At the largest ``fluid'' scales, $k_\perp\rho_i \lesssim 1$, linear phase mixing drives a velocity-space cascade in the Hermite moments along the magnetic-field direction, with a power law close to $m_\|^{-1/2}$. A dual phase-space cascade in ($k_\perp$,$m_\perp$) develops at scales below the ion gyroradius, $k_\perp\rho_i\gtrsim 1$, with $m_\perp \propto k^2_\perp$ likely due to nonlinear phase mixing \citep{SchekochihinPPCF2008,SchekochihinAPJS2009}. The resulting Hermite spectrum is steeper than our predictions based on gyrokinetic theory, but is close to the $m_\perp^{-2}$ spectrum predicted for magnetized turbulence in a Vlasov plasma \citep{ServidioPRL2017}. This steepness could be due to effects ordered out of the gyrokinetic theory, such as finite Larmor frequency~\citep[see, e.g.,][]{BrunoTrenchiAPJL2014,TelloniAPJ2015}, and/or due to multiscale dissipation or non-local couplings in phase space \citep[e.g.,][]{HatchJPP2014,CerriPOP2014a,PassotAPJL2015,TeacaNJP2017}. 

These results constitute the first evidence that both linear and nonlinear phase mixing are at play in magneto-kinetic plasma turbulence. Because the resulting spectral scaling laws are largely consistent with theories accounting for turbulent intermittency, we expect the consequent dissipation to be associated with intermittent structures such as current sheets and coherent magnetic structures. This, alongside the fact that reconnection seems to enhance/trigger the energy transfer below ion kinetic scales \citep{CerriCalifanoNJP2017,FranciAPJL2017}, may explain why most of the kinetic activity, energy conversion, and dissipation in solar-wind turbulence and in kinetic simulations is concentrated within or in the vicinity of ion-scale current sheets \citep[e.g.,][]{ServidioPRL2012,ServidioJPP2015,OsmanPRL2012,OsmanPRL2012b,WuAPJL2013,KarimabadiPOP2013,TenBargeAPJL2013,ChasapisAPJL2015,BanonNavarroPRL2016,WanPRL2015,WanPOP2016,YangPOP2017}. 
Although the present model neglects possible contributions from electron kinetics at sub-ion-Larmor scales \citep[e.g.,][]{ToldPRL2015}, we believe these ion-phase-space cascades constitute an important pathway to turbulent dissipation in collisionless plasmas. 

\acknowledgments

The authors thank Lev Arzamasskiy, Yuri Cavecchi, and especially Alex Schekochihin for valuable conversations, as well as the anonymous referee for a prompt and constructive report. 
S.S.C.~and F.C.~acknowledge Carlo Cavazzoni (CINECA, Italy) for essential contributions to the parallelization and optimization of the HVM code.
The simulation was performed at CINECA (Italy) under the ISCRA initiative (grant HP10BEANCY). S.S.C.~and M.W.K.~were supported by the National Aeronautics and Space Administration under Grant No.~NNX16AK09G issued through the Heliophysics Supporting Research Program. 


\end{document}